\documentstyle[twocolumn,pre,aps,psfig]{revtex}
\begin{document}
\draft
\title{Ising cubes with enhanced surface couplings}
\author{M. Pleimling and W. Selke}
\address{Institut f\"ur Theoretische Physik B, Technische Hochschule,
D--52056 Aachen, Germany}
 
\maketitle
 
\begin{abstract}
Using Monte Carlo techniques, Ising cubes with ferromagnetic
nearest--neighbor interactions and enhanced couplings
between surface spins are studied. In  
particular, at the surface transition, the corner magnetization 
shows non-universal, coupling-dependent 
critical behavior in the thermodynamic limit. Results on
the critical exponent of the corner magnetization are compared to
previous findings on two-dimensional Ising models
with three intersecting defect lines.\\

\end{abstract}


\section{Introduction}
 
In the thermodynamic limit, critical phenomena may occur not only
in the bulk of a system, but also
at its surfaces, edges and corners. To be specific, let us
consider Ising magnets with short-range interactions. Then, there are two
typical scenarios: (a) bulk, $m_b$, surface, $m_1$, edge, $m_2$,
and corner, $m_3$, magnetizations may order at the same temperature ("ordinary
transition"), but with different power--laws, and (b) surface, edge
and corner magnetizations may order simultaneously
first ("surface transition") due to enhanced, strong surface
couplings, followed by ordering of the bulk magnetization at the
lower bulk transition temperature ("extraordinary transition"). Surface
singularities at ordinary and surface
transitions have been studied extensively, theoretically 
\cite{binder,diehl} as well as experimentally \cite{dosch}. Most
of the rather few studies on edge critical 
behavior dealt with the ordinary
transition \cite{cardy,gut,peschel,ps2}. Only very recently, edge
criticality both at the surface transition \cite{ps3} and at the normal
transition \cite{han} has been investigated.

Similarly, corner criticality in three-dimensional 
Ising systems has been analysed, to our knowledge, only at the
ordinary transition, applying mean-field
theory \cite{igloi} and Monte Carlo simulations \cite{ps2}. However, that
case deserves to be studied at the surface transition as well for various
reasons. For magnetic properties of nanostructured materials, corners
are expected to play an important role \cite{meri,agu}. In
addition, magnetic couplings may be enhanced at
surfaces, especially at step-edges and corners \cite{kirsch}. Last, but
not least, the problem is of genuine theoretical interest. At 
the surface transition, the critical fluctuations
are essentially two--dimensional. Edges are local perturbations
acting then like defect lines in two-dimensional Ising models leading
to interesting non--universal critical
phenomena \cite{igloi,ps3}. Accordingly, corners, say, of an Ising
cube may be interpreted as intersection points of three defect
lines. At such points, one also expects intriguing non-universal behavior of
local quantities, such as the corner magnetization, following
exact analytical work on two--dimensional Ising systems 
with intersecting defect lines \cite{henkel}.\\
 
\section{Model, method and results}
 
We study nearest--neighbor Ising models on simple cubic lattices
with $L \times M \times N$ spins (usually we shall consider
Ising cubes, i.e. $L= M= N$) and ferromagnetic interactions. The
Hamiltonian may be written in the form

\begin{eqnarray}
{\cal H} & = & -\sum\limits_{bulk} J_b S_{xyz} S_{x'y'z'}
 -\sum\limits_{surface} J_s S_{xyz} S_{x'y'z'} \nonumber \\
& & -\sum\limits_{edge-surface} J_{es} S_{xyz} S_{x'y'z'}
 -\sum\limits_{edge} J_e S_{xyz} S_{x'y'z'}
\end{eqnarray}
where the sums run over bonds between neighboring
spins, $S_{xyz}= \pm 1$, with coupling constants to be specified
below; $x (y, z)$ going from 1 to $L (M, N)$. Free boundary conditions
hold for the spins in the surface planes. The
pairs of neighboring spins in the Hamiltonian (1) 
are located either on edge sites with the edge
coupling $J_e$, on edge and surface sites coupled by $J_{es}$, on
surface sites with the interaction $J_s$, or on sites with at least
one of the spins in the interior of the
system interacting with the bulk coupling $J_b$. We refrained from
assigning another coupling strength to pairs of spins on corner
and edge sites, which is taken to be equal to $J_e$.

To study the behavior at the surface transition, $T_s$, we
chose $J_s= 2 J_b$, where $k_BT_s/J_b \approx 4.975$ \cite{ps3}, while
the bulk transition, $T_c$, occurs
at $k_BT_c/J_b \approx 4.5115$ \cite{landau,tal}. The effect
of the edge couplings was studied by considering the three cases
(i) of equal surface couplings, i.e. $J_e= J_{es}= J_s$, (ii) of
reduced edge couplings, especially $J_{es}= J_s$, $J_e= J_b$, and
(iii) of reduced edge--surface couplings, especially $J_e= J_s$, $J_{es}= J_b$.

The size of the Ising cubes wih $L^3$ spins ranged from $L= 5$ to
$L= 80$. In the Monte Carlo simulations, we used the efficient
single--cluster--flip algorithm. Thermal averages were
obtained from an ensemble
of at least $10^2$ realizations, using different random numbers. In
each realization, several $10^4$ clusters were taken into account, after
equilibration. 

The quantity of main interest is the corner magnetization, or
more general, the local magnetization $m_l(x,y,z)$ at
site $(xyz)$. $m_l(x,y,z)$ may be defined by the correlation function 
\begin{equation}
 m_l(x,y,z)= \sqrt{< S_{xyz} S_{x'y'z'}>}
\end{equation}
where $(xyz)$ and $(x'y'z')$ are topologically equivalent sites
with maximal separation distance; brackets denote the thermal average. In
the thermodynamic limit, $L \longrightarrow \infty$, one
recovers the standard definition of the local magnetization
$m(x,y,z)= <S_{xyz}>$. $m_l(x,y,z)$ approaches closely
$m(x,y,z)$ provided the separation distance between the two equivalent
spins is large compared to the correlation length. Certainly, finite--size
effects are most severe near criticality, as usual. The deviation
of $m_l$ from $m$ may be monitored by varying the size of the cubes, $L$, and
by considering the correlation function between spins on 
equivalent sites with different separation distances (for
instance, two corner spins may be connected either by an edge, by
a surface diagonal, or by the bulk diagonal).  

%In addition, we recorded the total energy per spin, $E$, and the
%specific heat, $C$.

In Fig. 1 the intriguing profile of the local magnetization
at the surface, $m_l(x,y,1)$, of a $40^3$ Ising cube with
equal surface couplings, case (i), is depicted, at 
$k_BT/J_b= 4.9$, i.e. $T \approx 0.985 T_s$. The non--monotonic behavior
along pathes from the edges or corners towards the center of the
surface reflects the influence of bulk spins, as has been
discussed before \cite{ps3}. Crossover to monotonicity of
the profile, with the largest magnetization at the corners, is
expected to occur even closer to $T_s$. At lower temperatures, roughly
$T < T_c$, the profile is monotonic as well, with the smallest
magnetization at the corners due to the different coordination
numbers at corners, edges, and surfaces.

Near the surface transition, $T_s$, the corner
magnetization, say, $m_3= m_l(1,1,1)$, is expected to vanish, in
the thermodynamic limit, as
$m_3 \propto t^{\beta _3}$, where $t$ is the reduced temperature
$t= |T-T_s|/T_s$. To estimate $\beta_3$, we consider the effective
exponent \cite{ps2,ps3,ps1} $\beta_{eff}(t) = d \ln m_3/ d \ln t$. When
analysing the Monte Carlo data, the derivative is replaced by
a difference at discrete temperatures. As $t \longrightarrow 0$, $\beta_{eff}$
approaches $\beta_3$, provided finite--size
effects can be neglected.

The temperature dependence of the effective exponent $\beta_{eff}$
for the three different
sets of couplings, (i), (ii), and (iii), is shown in Fig. 2, displaying
only data which were checked to be unaffected by finite--size effects. Error
bars stem from the ensemble averaging performed to determine $m_3$. The
resulting estimates for the asymptotic critical exponent
$\beta_3$ are (i) $0.06 \pm 0.01$ at
$J_e= J_{es}= J_s= 2 J_b$, 
(ii) $0.14 \pm 0.015$ at $J_{es}= J_s= 2J_b$
and $J_e= J_b$, and 
(iii) $0.26 \pm 0.02$ at $J_e= J_s= 2 J_b$
and $J_{es}= J_b$.
In addition, we estimated $\beta_3$ at the ordinary
transition, $J_e= J_{es}= J_s= J_b$ to be $\beta_3= 1.77 \pm 0.05$, confirming
and refining our previous estimate based on computing the corner
magnetization from metastable states \cite{ps2}.-- Error bars are 
inferred from "reasonable" extrapolations of the effective exponent, see
Fig. 2.

To explain the Monte Carlo findings on $\beta_3$, note that the critical
fluctuations at the surface transition are essentially two--dimensional
and that corners are intersection points of the edges. Now, as had
been shown before \cite{ps3}, at the surface transition edges act as
ladder-- or chain--type defect lines \cite{fife,bariev,mccoy,igloi}. The
critical exponent $\beta_2$ of the edge
magnetization is non--universal (being non--trivial even in
case (i) of equal surface couplings, due to the coupling to
bulk spins), varying with the edge, $J_e$, and
edge--surface, $J_{es}$, couplings\cite{ps3}. To a given
set of interactions $J_e$ and  $J_{es}$, one
may assign roughly an effective defect coupling of
ladder-- or chain--type, $J_d^{eff}$, yielding
the same critical exponent for the defect magnetization in the
two--dimensional Ising model, $\beta_l$, and for the edge
magnetization at the surface transition of the three--dimensional
Ising model, $\beta_2$. Specificly, for
a ladder--type defect, the critical exponent of the
magnetization in the ladder rows, $\beta_l$, is given by \cite{bariev,mccoy}
\begin{equation}
 \beta_l = 2 \arctan^2(\kappa_l^{-1})/\pi^2
\end{equation}
with $\kappa_l= \tanh (J_l/(k_B T_{2d}))/ \tanh (J/(k_B T_{2d}))$, where
$T_{2d}$ is the transition temperature. Comparing
$\beta_2$ and $\beta_l$, one may interpret the defect coupling
$J_l$ of the two--dimensional model as the desired
effective coupling $J_d^{eff}$ ($J$ is the coupling constant away
from the defect line, corresponding to $J_s$ in the three--dimensional
systems).

Following this analogy, the critical exponent of the corner 
magnetization, $\beta_3$, can be related to that of the magnetization at the
intersection of three defect lines in the two--dimensional
Ising model, $\beta_i$, with effective defect couplings, $J_d^{eff}$.
 Indeed, in the two--dimensional 
Ising model, the value of $\beta_i$ has been
calculated exactly for three intersecting ladder defects
by Henkel et al. \cite{henkel}, showing
a non--universal behavior, with $\beta_i$ depending on
the strength of the defect couplings, $J_l$. If
those couplings are weaker than in the
rest of the system, then $\beta_i$ will increase with
decreasing $J_l (< J)$, $\beta_i > 1/8$, 1/8 being
the well--known Onsager value in the isotropic two--dimensional
Ising model. In turn, if the defect
couplings get stronger, then $\beta_i$ will get smaller. The
concrete expression for $\beta_i$ is quite lengthy \cite{henkel}
and will not be reproduced here, but
it can be evaluated in a straightforward way.
 
The effective ladder-type defect couplings $J_d^{eff}$ in the three cases we
considered are (i) $J_d^{eff} \approx 1.22 J_s$ corresponding to 
$\beta_2 \approx 0.095$ \cite{ps3} in the case 
of equal surface couplings, i.e. an effective enhancement of the
couplings at the edges due to the influence of bulk
spins, (ii) $J_d^{eff} \approx 0.99 J_s$ corresponding to $\beta_2 \approx 0.127$ \cite{ps3} for
weakened edge couplings, i.e. the
enhancement is now approximately compensated by the weakening of $J_e$, and 
(iii) $J_d^{eff} \approx 0.74 J_s$ corresponding
to $\beta_2 \approx 0.176$ \cite{ps3} for
weakened edge--surface couplings, overcompensating
the enhancement by the reduction in $J_{es}$ (note that 
the values of $\beta_2$ differ significantly from those
of $\beta_3$). Using
these estimates of $J_d^{eff}$, one 
obtains from the exact expression \cite{henkel} for the two--dimensional Ising
model with three intersecting ladder defects of those strengths the
following values for $\beta_i$  
(i) 0.082 , (ii) 0.128, and (iii) 0.21, in
satisfactory agreement with the Monte Carlo findings on 
$\beta_3$. Of course, a more refined analysis had to
take into account, e.g., the rather
complicated (see also the
non--monotonic profile in Fig. 1) nature of the edge
as a simultaneously ladder-- and chain--type
defect line as well as the effect of the bulk spin
next to the corner on the corner magnetization. Indeed, the good agreement 
between $\beta_3$ and $\beta_i$ in case (ii) may be related
to the fact that the chain--like character is rather weak in
that situation. The bulk spin is
expected to strengthen the effective coupling at the corner
especially in case (i), giving rise to the reduction
in $\beta_3$ as compared to $\beta_i$.      

Certainly, bulk properties will become critical only at the
extraordinary transition, at $k_BT_c/J_b \approx 4.5115$. For instance, the
specific heat $C$ is expected to diverge there, in
the thermodynamic limit. For finite, $L^3$ Ising cubes, one
observes that a maximum in $C$ near $T_c$ shows up only
for systems with at least a few thousands spins, getting more
pronounced as the system size increases (we
studied case (i) with equal surface couplings and $J_s= 2J_b$). On
the other hand, the
maximum in $C$ near the surface transition, $T_s$, dominates for
small cubes, becoming more and more suppressed as one increases
the size, $L$. For cubes of moderate size, say $15 < L < 60$, the
temperature dependence of the specific heat is characterised  
by an easily detectable two--peak structure, with maxima close
to $T_c$ and $T_s$.-The height of the two peaks may be easily
varied by replacing the Ising cubes by slabs.

In summary, the corner magnetization at the surface transition of Ising cubes
has been found to display non-universal critical behavior, with
the critical exponent $\beta_3$ of the corner
magnetization (being distinct from
the corresponding edge exponent $\beta_2$) depending
on the strength of the edge and edge--surface couplings. The
concrete value of $\beta_3$ may be approximated rather well from
the exactly known value of the critical exponent of the
magnetization at the intersection point of three defect lines
in the two--dimensional Ising model by estimating effective
defect couplings from the edge critical behavior.

\acknowledgments
It is a pleasure to thank M. Henkel and I. Peschel for very helpful
suggestions and discussions.

\begin{figure}
\centerline{\psfig{figure=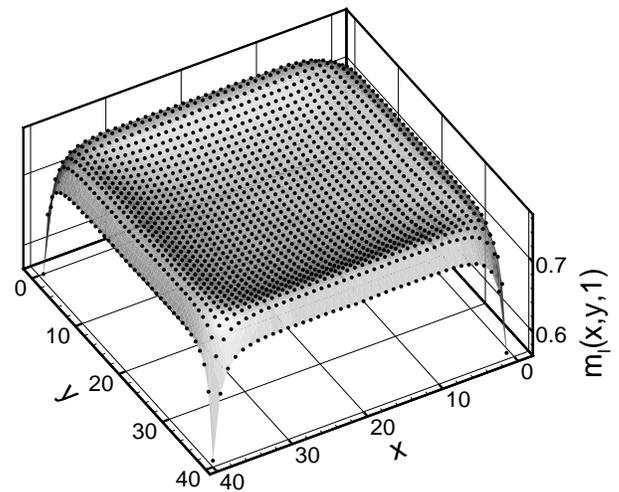,width=9.0cm}}
\caption{Simulated profile of the local
magnetization at the surface, $m_l(x,y,1)$, for an Ising
model of $40^3$ spins with equal surface
couplings, $J_e= J_{es}= J_s= 2 J_b$, at $k_BT/J_s= 4.9$.}
\label{fig1}
\end{figure}
 
\begin{figure}
\centerline{\psfig{figure=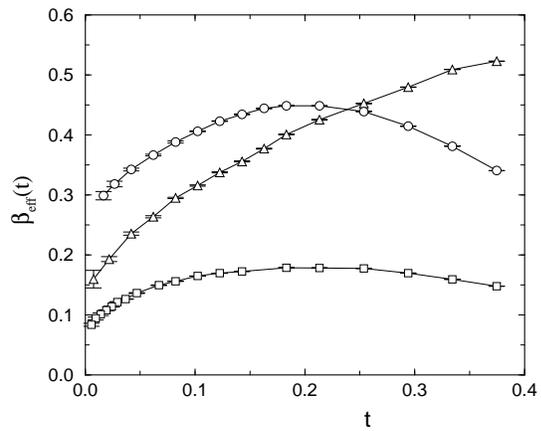,width=7.0cm,angle=270}}
\caption{Effective exponent $\beta_{eff}$ versus 
reduced temperature $t$ for 
(i) $J_e= J_{es}= J_s$ (squares) (ii) $J_{es}= J_s$, $J_e= J_b$ (triangles)
and (iii) $J_e= J_s$, $J_{es}= J_b$ (circles). Ising
cubes with up to $80^3$ spins were simulated, circumventing finite--size
effects.}
\label{fig2}
\end{figure}

\end{document}